\documentclass{article}
\usepackage{spconf,amsmath,graphicx,amsfonts}
\usepackage{multirow,makecell,hyperref}
\hypersetup{colorlinks,citecolor=black,filecolor=black,linkcolor=black,urlcolor=black}
\usepackage{enumitem}
\setlist{nosep, leftmargin=14pt}

\usepackage{mwe} 

\title{Self-Supervised Few-Shot Learning for Ischemic Stroke Lesion Segmentation}
 \name{Luca Tomasetti$^1$ \quad Stine Hansen$^2$ \quad Mahdieh Khanmohammadi$^1$ \quad Kjersti Engan$^1$} 
 \address{
  \textit{Liv Jorunn H\o{}llesli$^{1,3}$} \quad \textit{Kathinka D{\ae}hli Kurz$^{1,3}$} \quad \textit{Michael Kampffmeyer$^2$} \\
 $^{1}$ Department of Electrical Engineering and Computer Science, University of Stavanger, Norway \\
     $^{2}$ Department of Physics and Technology, UiT The Arctic University of Norway, Norway \\
     $^{3}$ Department of Radiology, Stavanger University Hospital, Norway }
\begin{document}
\ninept
\maketitle
\vspace{-0.2cm}
\begin{abstract}
Precise ischemic lesion segmentation plays an essential role in improving diagnosis and treatment planning for ischemic stroke, one of the prevalent diseases with the highest mortality rate.
While numerous deep neural network approaches have recently been proposed to tackle this problem, these methods require large amounts of annotated regions during training, which can be impractical in the medical domain where annotated data is scarce.
As a remedy, we present a prototypical few-shot segmentation approach for ischemic lesion segmentation using only one annotated sample during training.
The proposed approach leverages a novel self-supervised training mechanism that is tailored to the task of ischemic stroke lesion segmentation by exploiting color-coded parametric maps generated from Computed Tomography Perfusion scans. 
We illustrate the benefits of our proposed training mechanism, leading to considerable improvements in performance in the few-shot setting.
Given a single annotated patient, an average Dice score of 0.58 is achieved for the segmentation of ischemic lesions.
\end{abstract}
\begin{keywords}
Acute Ischemic Stroke, Few-Shot Segmentation, Self-Supervision, Supervoxels
\end{keywords}
\vspace{-0.2cm}
\section{Introduction}\label{sec:intro}
\vspace{-0.2cm}

Globally, neurological disorders are the leading cause of disability-adjusted life years and the second leading cause of death~\cite{world2022optimizing}. Cerebral stroke is a major contributor to this burden, being the number one cause of neurological disability and the third-leading cause of death and disability combined~\cite{feigin2019global,feigin2021global}. 
In response, the World Health Organisation has identified the optimization of brain health as a fundamental step to ensure human health and well-being \cite{world2022optimizing}.
Towards this goal, we focus on understanding and segmenting ischemic brain tissue in patients suspected of acute ischemic stroke (AIS), which comprises the majority of strokes \cite{katan2018global}.

\begin{figure}[tb]
\centering
\centerline{\includegraphics[width=\linewidth]{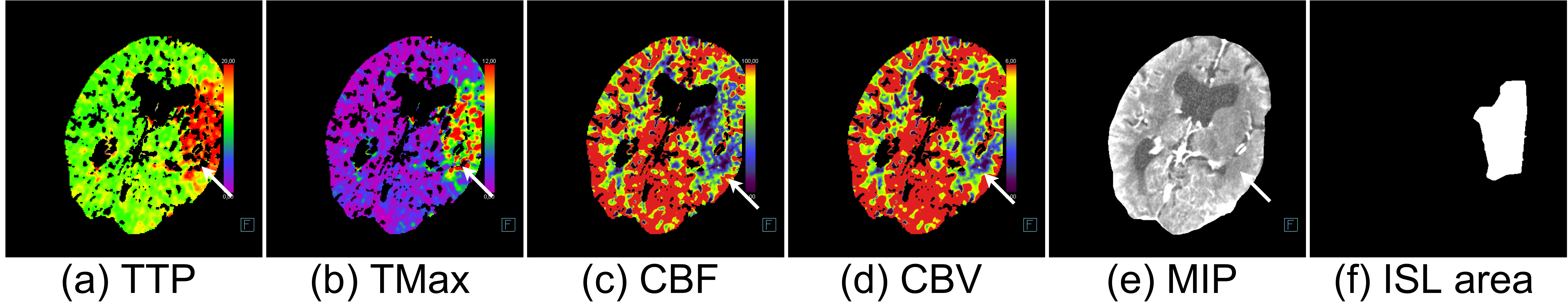}}
\vspace{-0.3cm}
\caption{Parametric maps (a-e) of a single brain slice plus the labeled ISL area (f). In this patient, there is an ischemic area in the vascular territory on the left middle cerebral artery (pointed by a white arrow).}
\vspace{-0.5cm}
\label{fig:pms}
\end{figure}

Computed Tomography (CT) Perfusion (CTP) is often used as the primary diagnostic modality in the initial stages of an AIS \cite{european2008guidelines}. 
A CTP study is a 4D examination of an area of the brain where an ischemic stroke lesion (ISL) is suspected. 
The study collects a series of 3D scans over the same portion of the brain during injection of an iodinated contrast agent over time.
Based on the brain tissue changes over time, color-coded parametric maps (PMs) are calculated.
These PMs highlight the possible variations in the brain tissue, allowing neuroradiologists to diagnose and plan treatments rapidly \cite{kurz2016radiological,tomasetti2021machine}.
In particular, we consider the cerebral blood flow (CBF), cerebral blood volume (CBV), time-to-peak (TTP), and time-to-maximum (TMax) PMs and maximum intensity projection (MIP), which are all highly relevant indicators for ISL segmentation~\cite{kurz2016radiological}.
Fig. \ref{fig:pms} shows a set of PMs and MIP for an image slice and the corresponding labeled ISL.

Numerous applications have been explored to tackle this neurological disorder, traditionally using thresholding of different PMs~\cite{bathla2019achieving,bivard2014defining}, but with lacking consensus on which PMs and threshold combination. 
Newer methods are relying on machine learning (ML) approaches \cite{tomasetti2021machine,qiu2020automated} or deep neural network (DNN) architectures \cite{amador2022predicting,tomasetti2022multi} and have generated promising results for segmenting ISLs.
Nevertheless, all these architectures are supervised learning methods and require large amounts of training examples manually annotated by expert neuroradiologists.
This can be problematic in the medical domain, where annotated training data is often scarce due to the time-consuming nature of the labeling process.
Moreover, manual annotations might only give a coarse representation of the ISL, excluding all the minor details in the area, which can lead to imperfect predictions if used during model training. 

\begin{figure*}[!ht]
\centering
\centerline{\includegraphics[width=.94\linewidth]{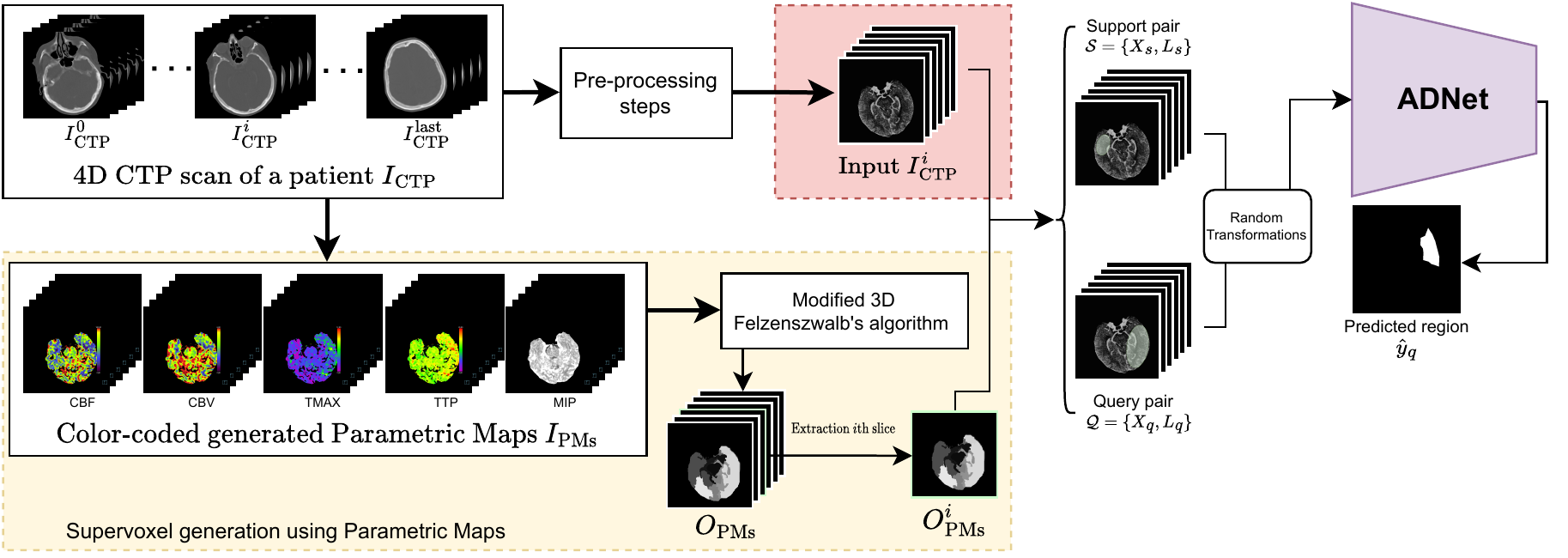}}
\vspace{-0.3cm}
\caption{General overview of the various steps involved in the method. Supervoxel regions are generated from the color-coded PMs with a modified 3D version of Felzenszwalb's image segmentation algorithm~\cite{felzenszwalb2004efficient}. A series of pre-processing steps (details in Sec. \ref{sec:pre}) are implemented to extract brain tissue. Two random slices containing the same supervoxel are then sampled to act as the support and query slice, with the supervoxel region acting as the pseudolabel for the self-supervised task.
Random transformations are then applied to either the support or query volume to provide the support-query pair that is used as input to train the ADNet model.}
\vspace{-0.5cm}
\label{fig:over}
\end{figure*}

A promising alternative, that has been gaining momentum in the last years, is self-supervised few-shot segmentation (FSS) \cite{ouyang2020self,hansen2022anomaly} where DNN algorithms only require a few labeled training images.
This can be crucial in the medical domain for reducing the need for annotated data and enabling the models to also learn rare cases for which annotated data can be scarce.
These models are trained and tested in episodes where a few labeled support images are used to guide the segmentation of the unlabeled query images.  
To bypass the need of manually annotating the training data, the training episodes are constructed in a self-supervised manner by leveraging automatically generated superpixel/supervoxel-based pseudolabels.
Then, during inference, only a few manually labeled support images are required to segment the unlabeled query images.

Ouyang et al. \cite{ouyang2020self} introduced a novel FSS framework for detecting organs in medical images using superpixel-based pseudolabels rather than manual annotations during training.
Hansen et al. \cite{hansen2022anomaly} further extended the self-supervision task to leverage 3D information via supervoxels and proposed a novel anomaly detection-inspired approach for prototypical FSS to increase the model's robustness to the heterogeneous background, resulting in the current state-of-the-art self-supervised FSS model.
During each training episode, two random slices containing the same random supervoxel are sampled to act as the support and query slice, with the supervoxel region acting as the pseudolabel for the self-supervised task.
Random transformations are then applied to the support or query image-pseudolabel pair and the objective is to segment the query's pseudolabel using as reference the support pair.
This gives the model a procedure to train unsupervised, i.e., without using labeled data.

These recent methods rely on the extraction of superpixels/supervoxels from the raw input images, which works well when the task is to segment classes that are relatively homogeneous and distinctly different from the background (e.g., liver or kidneys in abdominal CT scans) and where regions of interest thus are likely to be grouped in supervoxels. However, their performance decreases considerably when the class to segment is not distinctly different from the background, as is the case in ISL segmentation.
For this reason, we propose an extension of the approach developed by Hansen et al. \cite{hansen2022anomaly} that is particularly suited for the task of ISL segmentation.
Unlike Hansen et al. \cite{hansen2022anomaly}, we do not generate supervoxels directly from the raw 4D CTP study, but instead we leverage domain knowledge generating supervoxels as targets for the self-supervised learning task from the PMs.
We argue that while supervoxels generated from raw 4D CTP studies cannot correctly describe the boundaries inside the brain tissue, supervoxels extracted from the PMs can capture informative sub-regions, thus resulting in improved training targets. To the best of our knowledge, this is the first study exploiting self-supervised mechanisms for ISL segmentation.

The main contributions of this work can be summarized as follows:
\begin{enumerate}
    \item We propose a novel self-supervised mechanism specifically tailored for ISL segmentation that leverages supervoxels extracted from a set of stacked PMs.
    \item Exploiting the self-supervised mechanism, we propose a prototypical FSS framework for ISL segmentation and demonstrate the benefit of leveraging PMs for self-supervision.
\end{enumerate}

\vspace{-.4cm}
\section{Data Material}
\vspace{-0.2cm}
We have analyzed 152 CTP scans from distinct patients acquired with a Siemens CT scanner between January 2014 and August 2020.
Patients were divided into two groups based on the level of vessel occlusion: 77 patients had a large vessel occlusion (LVO) and 60 patients had a non-large vessel occlusion (Non-LVO). In addition, we included 15 patients admitted with stroke-like symptoms, but after diagnostic work-up were determined not to have suffered from an ischemic stroke, defined as without ischemic stroke (WIS). 
A split of 89/30/33 patients was conducted for training, validation, and testing. 
The CTP scans and the relative PMs were obtained using the ``syngo.via" software from Siemens Healthineers (Siemens Medical Solutions, USA).
Two expert neuroradiologists with 17 and 4.5 years of clinical experience manually annotated the ischemic areas using an in-house software.
The annotations were delineated by examining the entire set of PMs (CBV, CBF, TTP, TMax) and MIP  generated from a CTP study. 
Moreover, they used as assistance follow-up examination, including Magnetic Resonance Imaging (MRI) for most patients, performed within 1 to 3 days after the initial CT examination to assess the final infarction areas.

\vspace{-.4cm}
\subsection{Pre-processing steps}\label{sec:pre}
\vspace{-.2cm}
Each CTP study underwent a series of pre-processing steps to extrapolate brain tissue from the raw scans and to enhance the contrast. 
The pre-processing steps can be summarized as follow: 
\begin{enumerate}
    \item Each study is converted into Hounsfield unit (HU) values to have a known quantitative scale for describing radio-density.
    \item Co-registration of each slice is executed using the first slice as a frame of reference.
    \item Brain tissue is extracted using an algorithm by Najm et al. \cite{najm2019automated}. This algorithm was selected due to its public availability and proven efficiency for CT scans.
    \item Gamma correction and histogram equalization are performed after the brain's extraction for contrast enhancement.
\end{enumerate}

\vspace{-0.4cm}
\section{Method}
\vspace{-0.2cm}
In this work, we propose a PM-based self-supervised few-shot segmentation model for ISLs in patients suspected of AIS\footnote{Code is publicly available at \href{https://github.com/Biomedical-Data-Analysis-Laboratory/ADNet-for-AIS-segmentation}{https://github.com/Biomedical-Data-Analysis-Laboratory/ADNet-for-AIS-segmentation}}. The task is formulated as a binary segmentation, where the foreground corresponds to the area of interest. 
Instead of extracting pseudolabels from the raw input as in previous approaches~\cite{ouyang2020self,hansen2022anomaly}, which for ISLs will lead to sub-optimal performance due to the homogeneous nature of the brain region, we leverage domain-knowledge to improve the self-supervision task. 
In particular, we extract pseudolabels from the color-coded PMs that capture more informative sub-regions. 
After generating the PMs from the 4D CTP input data and extracting pseudolabels, we take inspiration from the ADNet model proposed in~\cite{hansen2022anomaly} to learn a binary classifier that can segment the CTP input based on one annotated image slice. Fig. \ref{fig:over} gives a general overview of all the steps involved in the proposed method.

\vspace{-.35cm}
\subsection{Parametric Map-Based Self-Supervision network}\label{pmsuper}
\vspace{-.2cm}
The proposed PM-based self-supervised network is an extension of the ADNet \cite{hansen2022anomaly} and is trained using supervoxel-based pseudolabels extracted from the unlabeled training images. 
A supervoxel can be considered a group of neighboring voxels in an image with a similar nature.
To obtain the supervoxels, we follow prior approaches in FSS~\cite{ouyang2020self, hansen2022anomaly} and use the Felzenszwalb's algorithm~\cite{felzenszwalb2004efficient}, as it provides more diverse supervoxels compared to alternative approaches such as Simple Linear Iterative Clustering (SLIC)~\cite{achanta2010slic}.
We leverage the 3D version of Felzenszwalb's algorithm to produce supervoxels from the 3D image volume\footnote{Code is publicly available at \href{https://github.com/sha168/Felzenszwalb-supervoxel-segmentation}{https://github.com/sha168/Felzenszwalb-supervoxel-segmentation}} and extend the Euclidean distance calculation within the algorithm to support that each voxel in the image is represented by a vector, thus accepting 4D input tensors $I \in \mathbb{R}^{(W \times H \times Z \times M)}$ and returning a 3D volume $O \in \mathbb{R}^{(W \times H \times Z)}$ with the corresponding segmented regions.
The four dimensions of the Felzenszwalb's input $I$ are defined as width ($W$), height ($H$), depth ($Z$), and an additional dimension ($M$), corresponding to the number of modalities \footnote{CBV, CBF, TTP, TMax, MIP in our proposed approach}, hence $M=5$; we named the Felzenszwalb's input $I_{\text{PMs}}$ and the corresponding output $O_{\text{PMs}}$.
A CTP scan after the pre-processing steps, which corresponds to the input of the proposed method, is called $I_{\text{CTP}} \in \mathbb{R}^{(W \times H \times Z \times M)}$, where the $M$ corresponds to the time dimension ($M=T=30$).
Let $X_s$ and $X_q$ represent the support and query 2D+time slices extracted from the same brain volume $I_{\text{CTP}}$, respectively. 
The algorithm relies on one parameter, $\rho$, which controls the supervoxel size. 
Experiments were performed to tune $\rho$ to improve the segmentation accuracy of the ischemic areas.
\begin{figure}[t]
\centering
\centerline{\includegraphics[width=.9\linewidth]{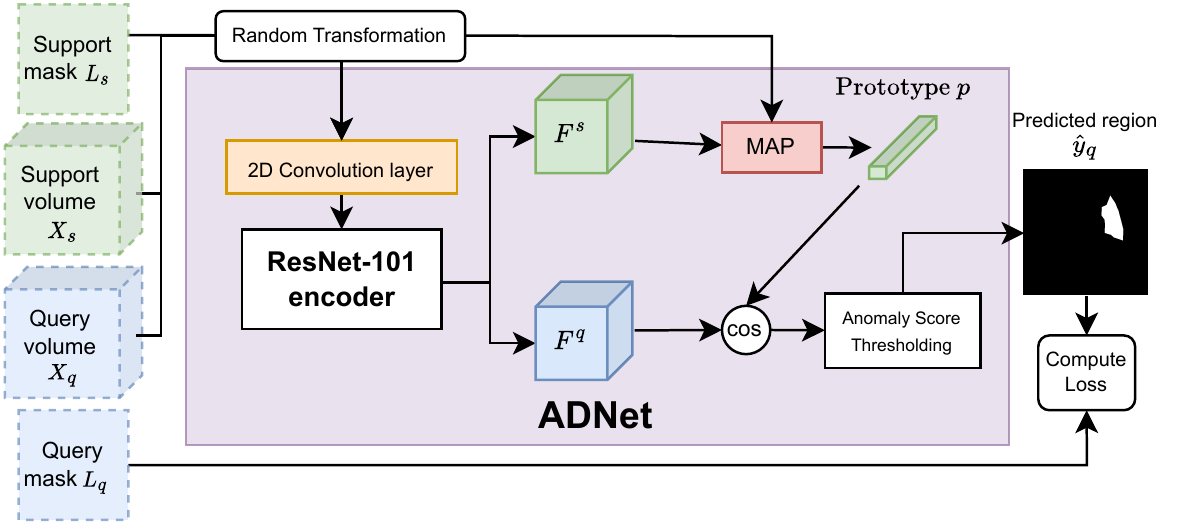}}
\vspace{-0.3cm}
\caption{Visual overview of the ADNet architecture \cite{hansen2022anomaly}. 
}
\vspace{-0.5cm}
\label{fig:adnet}
\end{figure}

An illustration of the ADNet architecture is shown in Fig. \ref{fig:adnet}. 
During training, the proposed model accepts support  $\mathcal{S} = \{X_s, L_s\}$ and query $\mathcal{Q} = \{X_q, L_q\}$ pairs as input, where $L_s$ and $L_q$ are the 2D binary label images for $X_s$ and $X_q$.
In particular, each training episode is constructed from one unlabeled CTP scan by first choosing a random supervoxel extracted from the PMs to act as the foreground class, and then samples two 2D image slices (+time) containing this supervoxel to act as support and query image ($X_s$ and $X_q$).
Training labels $L_s$ and $L_q$ are constructed from the supervoxel segmentations of the corresponding PMs: the selected supervoxel is foreground while the union of the remaining supervoxels acts as background class.  
Random transformations are applied to one of the support or query volumes, based on a 50\% probability, to encourage invariance to shape, and intensity differences \cite{ouyang2020self}.
$X_s$ and $X_q$ are fed to a shared feature encoder to extract high-level features $F_s$ and $F_q$, respectively.
The support foreground mask $L_s$ is used to perform masked average pooling (MAP) of the support features $F^s$ to compute a foreground prototype $p \in \mathbb{R}^d$, where $d$ is the dimension of the embedding space. 
To segment the query volume, the negative cosine similarity ($CS$) is computed between the prototype $p$ and the feature map $F_q$. 
The predicted foreground mask $\hat{y}_q$ is found by thresholding the computed $CS$ with a learned parameter $Th$ and the loss proposed in~\cite{hansen2022anomaly} is optimized given the query label $L_q$.
During inference phase, we follow the evaluation protocol of \cite{hansen2022anomaly} and sample one single 2D+time slice $X_s$ from a support patient. The selected slice is always the middle slice of the support volume to avoid limited or non-existing ischemic regions at the edges of the volumes. 
$L_s$ is the corresponding 2D slice with labeled ISL. The support pair $\{X_s, L_s\}$ is used to segment ISLs in the entire query patient one slice at a time.
For a detailed explanation of the steps involved in the training and inference phases, we refer the reader to \cite{hansen2022anomaly}. 

\vspace{-.45cm}
\section{Experiments}
\vspace{-.3cm}

\begin{table*}[ht!]
    \centering
    \resizebox{\linewidth}{!}{%
    \begin{tabular}{c|c|c||c|c|c||c|c|c||c|c|c|c}
    \hline
        \multirow{2}{*}{\textbf{Method}} & \multirow{2}{*}{\textbf{ Best $ \rho$}} & \multirow{2}{*}{\textbf{Dataset}} &  \multicolumn{3}{c||}{\textbf{DS} $\Uparrow$} & \multicolumn{3}{c||}{\textbf{MCC} $\Uparrow$} & \multicolumn{4}{c}{\textbf{$\Delta V$ (ml)} $\Downarrow$} \\ \cline{4-13}
        & & & LVO & Non-LVO & All & LVO & Non-LVO & All &LVO & Non-LVO & WIS & All  \\  \Xhline{3\arrayrulewidth}
        \hline
        \multicolumn{13}{c}{\textit{Self-Supervised Methods}} \\
        \hline
        \textit{CTP-Baseline} & 1000 & \multirow{3}{*}{CTP-LVO} & 0.50$\pm$0.14 & 0.07$\pm$0.03 & 0.43$\pm$0.12 & 0.48$\pm$0.14 & 0.10$\pm$0.05 & 0.43$\pm$0.12 & 72.6$\pm$15.8 & 85.4$\pm$26.5 & 212.1$\pm$63.0 & 89.5$\pm$15.4  \\ \cline{1-2} \cline{4-13}
        \textit{PMs-Baseline} & 1250 && 0.50$\pm$0.15 & 0.09$\pm$0.03 & 0.45$\pm$0.14 & 0.48$\pm$0.15 & 0.11$\pm$0.03 & 0.44$\pm$0.14 & 74.8$\pm$15.7 & 50.3$\pm$42.8 & 75.5$\pm$41.3 & 66.7$\pm$19.5  \\ \cline{1-2} \cline{4-13}
        Proposed & 1250 && \textbf{0.60$\pm$0.16} &\textbf{0.13$\pm$0.05} & \textbf{0.58$\pm$0.16} & \textbf{0.58$\pm$0.17} & \textbf{0.14$\pm$0.05} & \textbf{0.57$\pm$0.16} & \textbf{56.4$\pm$14.7} & \textbf{23.0$\pm$40.0} & \textbf{16.0$\pm$50.0} & \textbf{41.6$\pm$25.4}  \\  \hline \noalign{\vskip .05in} \hline       
        \textit{CTP-Baseline} & 1000 & \multirow{3}{*}{CTP-ALL} & 0.47$\pm$0.11 & 0.09$\pm$0.02 & 0.39$\pm$0.09 & 0.45$\pm$0.11 & \textbf{0.14$\pm$0.04} & 0.39$\pm$0.10 & 84.2$\pm$13.4 & 128.9$\pm$23.3 & 176.3$\pm$31.5 & 107.5$\pm$17.4 \\ \cline{1-2} \cline{4-13}
        \textit{PMs-Baseline} & 1750 && 0.50$\pm$0.13 & \textbf{0.10$\pm$0.05} & 0.45$\pm$0.12 & 0.48$\pm$0.14 & 0.14$\pm$0.07 & 0.43$\pm$0.13 & 86.3$\pm$15.1 & 70.5$\pm$34.1 & 219.6$\pm$57.9 & 93.2$\pm$15.6 \\ \cline{1-2} \cline{4-13}
        Proposed & 1500 && \textbf{0.55$\pm$0.10} & 0.06$\pm$0.03 & \textbf{0.51$\pm$0.10} & \textbf{0.54$\pm$0.10} & 0.07$\pm$0.03 & \textbf{0.50$\pm$0.10} & \textbf{60.4$\pm$10.9} & \textbf{32.3$\pm$6.8} & \textbf{13.8$\pm$12.0} & \textbf{46.8$\pm$9.8} \\ \hline \noalign{\vskip .05in} \hline
        \multicolumn{13}{c}{\textit{Supervised Method} (used only for comparison)}\\
        \hline
        Tomasetti et al.~\cite{tomasetti2022multi} & N.A. & CTP-ALL & 0.85$\pm$0.06 & 0.49$\pm$0.30 & 0.66$\pm$0.32 & 0.83$\pm$0.06 & 0.49$\pm$0.29 & 0.64$\pm$0.31 & 19.6$\pm$15.2 & 4.9$\pm$5.4 & 0.3$\pm$0.4 & 12.9$\pm$14.2 \\ \hline
    \end{tabular}}
    \vspace{-0.4cm}
    \caption{Mean (+ standard deviation) test results on the two datasets. All the methods (with the exception of Tomasetti et al.~\cite{tomasetti2022multi}) utilize a single labeled 2D+time slice during inference phase. Tomasetti et al.~\cite{tomasetti2022multi} uses a supervised approach and relies on the full labeled dataset for training and inference.
    Note that higher values are preferable for the DS and MCC ($\Uparrow$), while for $\Delta V$, lower values are better ($\Downarrow$). Best performance are highlighted in $\textbf{bold}$ (last row is only used for comparing supervised method and our self-supervised approaches).}
    \label{tab:exps}
\end{table*}

The proposed method is implemented to segment ISLs in patients suspected of AIS.
In the following, we illustrate the benefit of leveraging the domain knowledge in the form of color-coded PMs for supervoxel extraction, rather than the raw CTP. 
For this, we compare the proposed method to a baseline approach that directly applies ADNet to the raw CTP scans, called \textit{CTP-Baseline}. To further demonstrate the need to jointly leverage both the raw CTP scans and the PMs, we compare to another baseline that utilizes PMs both as input and for the pseudolabel generation, which we term \textit{PMs-Baseline}. 

\vspace{-0.5cm}\paragraph*{Baselines}
Our \textit{CTP-Baseline} model uses supervoxels generated from raw 4D CTP studies. For the Felzenszwalb's algorithm the 4D input is $I_{\text{CTP}}$. The same input is used in the model.
The \textit{PMs-Baseline} method utilizes supervoxels generated directly from a set of 3D PMs stacked over the fourth dimension (Sec. \ref{pmsuper}). 
The 4D input, for producing supervoxels and for the \textit{PMs-Baseline}, is $I_{\text{PMs}}$.

\vspace{-0.5cm}\paragraph*{Evaluation}
To demonstrate the applicability of our proposed method, we performed evaluations on the following datasets:
\begin{itemize}
    \item \textbf{CTP-ALL} comprises the full dataset consisting of all three groups involved in this study (152 subjects). 
     \item \textbf{CTP-LVO} contains only patients within the LVO group (77 subjects). Patients from this group are a clinically significant share of patients with AIS, considering the extension of the ischemic area and the grim natural course of the condition. This dataset is implemented to illustrate the variation effects during training.
\end{itemize}
Following standard practice in the literature \cite{ouyang2020self, hansen2022anomaly}, we evaluate the performance by measuring the Dice score (DS); furthermore, we measure the predictions with the Matthews Correlation Coefficient (MCC)  \cite{matthews1975comparison}, which has been proven to provide a better measurement for imbalanced datasets \cite{chicco2017ten}.
Moreover, we employ the absolute difference in volumes $\Delta V(V_y, V_{\hat{y}}) = |V_y - V_{\hat{y}}|$ between the predicted volume $V_{\hat{y}}$ and the manually annotated volume $V_y$. This metric is essential for the WIS group because of the lack of labeled areas in this group, which makes it impossible to rely on the DS or the MCC.

We evaluate the effect of leveraging supervoxels obtained from the PMs via a comparison of the performance of our proposed method and the baseline approaches. 
In order to select the best hyperparameter $\rho$ for the three models, we further performed a sensitivity study where we analyzed the parameter $\rho$ in relation to the aggregate DS and the $\Delta V$ from the set of validation patients.
These two-step experiments help create a fair comparison among the analyzed models.


\vspace{-.5cm}
\section{Results \& Discussion}
\vspace{-.3cm}
Table \ref{tab:exps} compares our proposed method with the baseline approaches for the test datasets split into the three groups (LVO, Non-LVO, WIS). 
Moreover, a comparison with the current state-of-the-art supervised method~\cite{tomasetti2022multi} is presented.
Results are given as the mean (+ standard deviation) of $22$ inference runs with a single slice from different patients as support.
The proposed method and baseline results are shown with the best-selected $\rho$ parameter chosen over a set of experiments on the validation set.
In Table \ref{tab:exps}, we can see that our proposed model outperforms the baseline approaches for both datasets.
The overall best performance is obtained by our proposed method using the \textbf{CTP-LVO} dataset for training, which indicates that the use of a single group of patients during the training phase gives a better understanding of the ischemic region.
\begin{figure}[t]
\vspace{-0.4cm}
\centering
\centerline{\includegraphics[width=.9\linewidth]{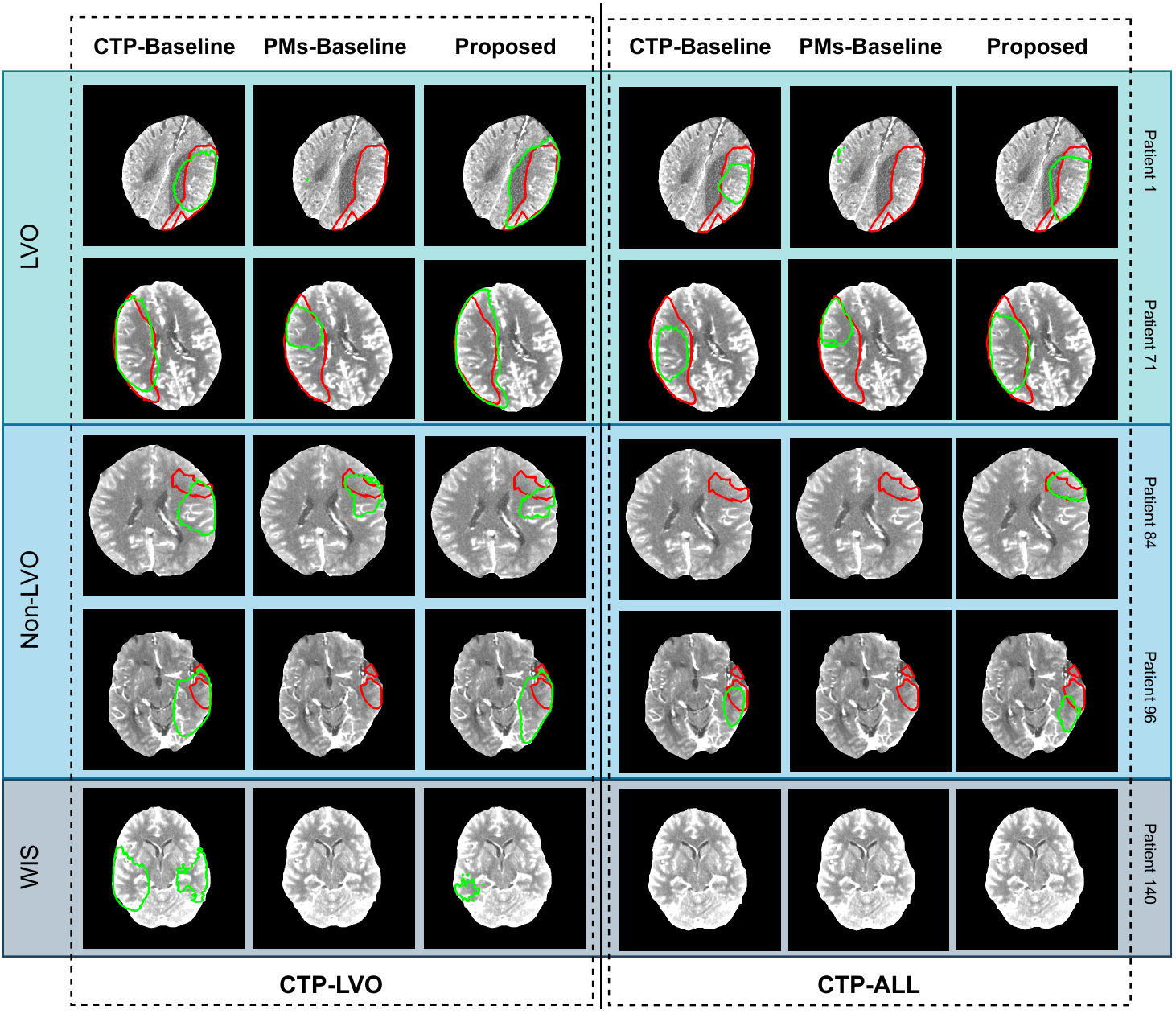}}
\vspace{-0.4cm}
\caption{Qualitative results of predicted ISL from random brain slices of different patients from the test set. Predictions are in green, while the ISL manually annotated ground truth are in red.
}
\vspace{-0.5cm}
\label{fig:res}
\end{figure}

The qualitative results in Fig. \ref{fig:res} support the quantitative results in Table \ref{tab:exps}.
Training the models with images from the \textbf{CTP-LVO} dataset results in satisfactory segmentations for the ISLs, especially for the LVO group.
With the adoption of the \textbf{CTP-ALL} as the training dataset, the results follow the correct structure of the ISL but with a tendency of under-segmentation in comparison with the method trained with the \textbf{CTP-LVO} dataset.
Outcomes from our proposed approaches (third and last columns in Fig. \ref{fig:res}) are not perfect, but they strongly reflect the area in the brain with an ISL, especially for patients in the LVO group.
The \textit{CTP-Baseline} model brings unnecessary under-segmentation (fourth column in Fig. \ref{fig:res}) or excessive over-segmentation (first column in Fig. \ref{fig:res}), depending on the dataset used for training. The \textit{PMs-Baseline} approach yields more stable segmentations across different support images.
Employing $I_{\text{PMs}}$ as input in creating the supervoxels and using them in combination with $I_{\text{CTP}}$ as input for the model provides a superior balance in the predicted segmentation regions; however, none of the methods are suitable for predicting ISL in Non-LVO patients. 
The meager nature of these areas makes it harder for the model to understand and adequately segment the ISL.
The proposed approach, trained with the \textbf{CTP-LVO} dataset, achieved the best metrics among all the analyzed groups.
This implies that the proposed PMs-Based Self-Supervision method is more suitable to segment ischemic regions and that there is a benefit to using PMs for supervoxel generation rather than relying on supervoxels extracted from raw CTP studies.
Although the performances are not as high as the supervised method~\cite{tomasetti2022multi}, which utilizes the entire labeled dataset during training, the achieved segmentation accuracy for the LVO group is still encouraging. This accuracy was attained using only a single labeled 2D+time slice, emphasizing the significance of employing domain knowledge in a limited labeled data setting

\begin{figure}[t]
\vspace{-0.6cm}
\centering
\centerline{\includegraphics[width=.85\linewidth]{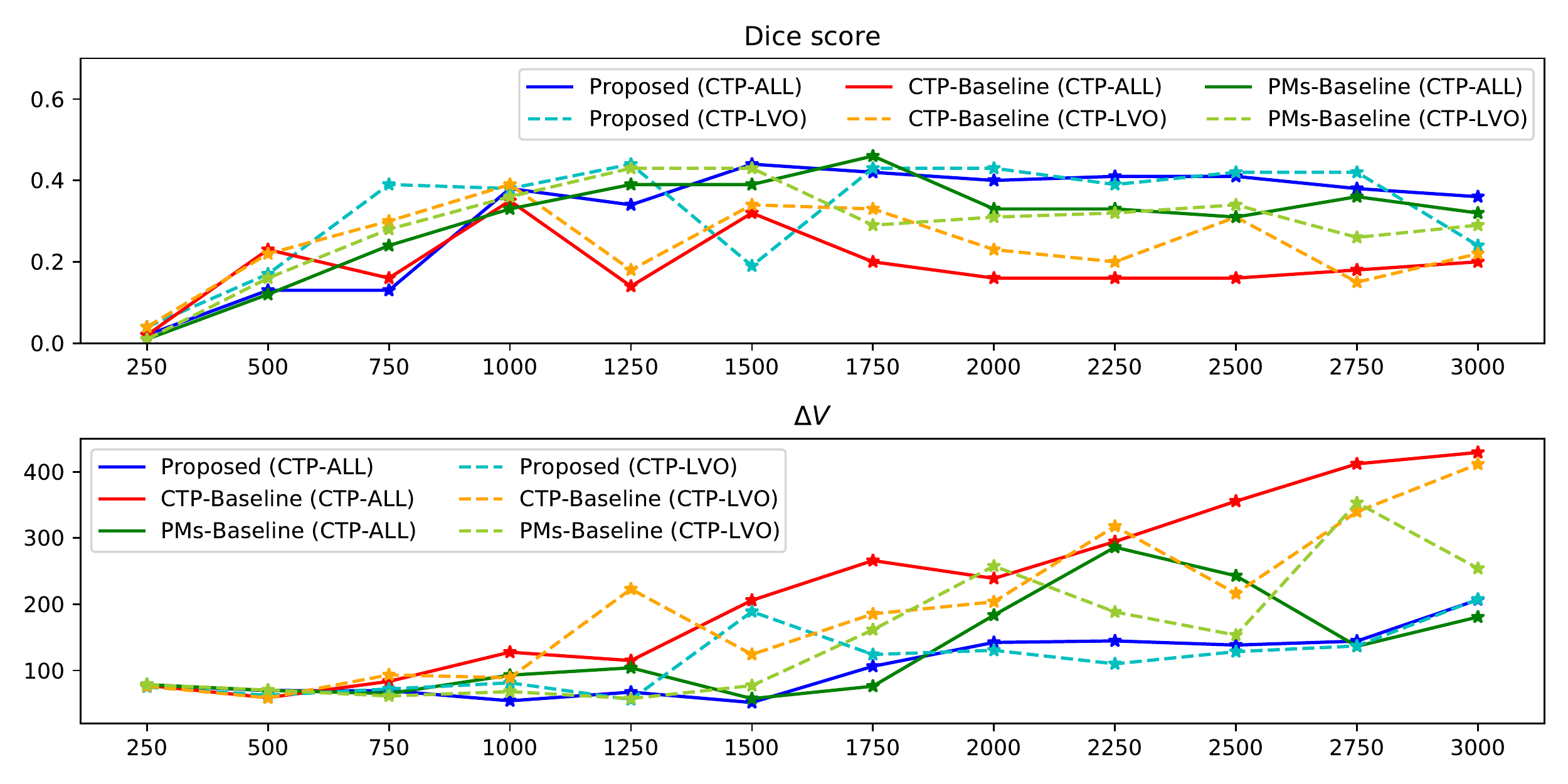}}
\vspace{-0.5cm}
\caption{Analysis of various $\rho$ parameters (x-axis) in relation to the DS results (top graph) and the $\Delta V$ (bottom graph) on the y-axis for the validation set for the two datasets used.
}
\vspace{-0.6cm}
\label{fig:rho}
\end{figure}

\vspace{-0.5cm}\paragraph*{Sensitivity Study}
The performance for a range of $\rho$ values is illustrated in Fig. \ref{fig:rho}.
As it is possible to evince from the graphs, the performance of the proposed method are relatively robust in the size of the supervoxels inside this range ($1000<\rho<2500$), while for the baseline, the range was $1000<\rho<1750$.
Outside this range, the performances are inferior for the DS, while the $\Delta V$ increase with the increment of the parameter $\rho$. 

\vspace{-.5cm}
\section{Conclusion}
\vspace{-.3cm}
We proposed a novel self-supervised mechanism to perform few-shot segmentation of ISLs and integrated it into the current state-of-the-art self-supervised FSS model ADNet. Instead of leveraging supervoxels directly generated from the raw CTP data for self-supervised training, which leads to sub-optimal performance, the proposed model leveraged domain knowledge by extracting supervoxels from parametric maps using a modified 3D version of Felzenszwalb's algorithm.
The present study demonstrated the benefit of leveraging domain knowledge, especially for patients affected by a large vessel occlusion. 
This study is a first step of exploring few-shot learning approaches for ISL segmentations.
Further research is still required to validate the results. 

\section{Compliance with ethical standards}
\label{sec:ethics}
This study was performed in line with the principles of the Declaration of Helsinki. Approval was granted by the Regional Ethic Committee Project 2012/1499.
      


\section{Acknowledgments}\label{sec:acknowledgments}
This work was supported by The Research Council of Norway (RCN), through its Centre for Research-based Innovation funding scheme [grant number 309439]; RCN FRIPRO [grant number 315029]; RCN IKTPLUSS [grant number 303514]; and the UiT Thematic Initiative.
The authors would like to thank the Neuroscience Research Group, SUS, under the supervision of prof. Martin W. Kurz for providing clinical patient information.






\bibliographystyle{IEEEbib}
\bibliography{refs}

\end{document}